# Threat Modeling Data Analysis in Socio-technical Systems

Tomasz Ostwald

Salient Works

Kirkland, WA, USA

tomasz@salientworks.com

## ABSTRACT
Our decision-making processes are becoming more data driven, based on data from multiple sources, of different types, processed by a variety of technologies. As technology becomes more relevant for decision processes, the more likely they are to be subjects of attacks aimed at disrupting their execution or changing their outcome. With the increasing complexity and dependencies on technical components, such attempts grow more sophisticated and their impact will be more severe. This is especially important in scenarios with shared goals, which had to be previously agreed to, or decisions with broad social impact. We need to think about our decisions-making and underlying data analysis processes in a systemic way to correctly evaluate benefits and risks of specific solutions and to design them to be resistant to attacks. To reach these goals, we can apply experiences from threat modeling analysis used in software security. We will need to adapt these practices to new types of threats, protecting different assets and operating in socio-technical systems. With these changes, threat modeling can become a foundation for implementing detailed technical, organizational or legal mitigations and making our decisions more reliable and trustworthy.

## 1.INTRODUCTION
Internet and mobile devices changed our interactions and communication patterns, and we can see how data analysis technologies are changing our decision-making practices and habits. New technologies become a natural foundation for our daily decisions that impact all aspects of our lives: personal, social, economic and political. They lead to new types of businesses, opportunities and relationships and can make our decision-making processes more effective, accurate and rational. This is especially important for decision situations with shared goals, which had to be agreed to, or decisions with broad social impact, as they usually require significant transparency, accountability and independence.

But we live in an imperfect world and we should never focus only on opportunities and benefits of new solutions. New technologies enable new scenarios that lead to new threats, which may require new mitigations. As technology becomes more useful and prevalent, decision processes based on data analysis will be targets of attacks aimed at disrupting their execution or changing their outcome. There are different technical, legal or organizational methods that can be used to mitigate such threats. But to implement them successfully, we need a systemic approach to understand and evaluate risks so that we could make informed decisions about design of our processes and dependencies on specific data sources or data processing components. We can base such analysis on threat modeling methodologies from software security, as they were specifically designed to look at complex systems from an attacker's point of view [1].

## 2.DECISION-MAKING BASED ON DATA ANALYSIS
The changes in our decision-making patterns are consequences of increasing availability of data, which were limited and had to be actively searched and acquired before the digital revolution. Currently we are usually much closer to a state of information overload than sensory deprivation as there is a lot of data that are automatically collected and easily available (at least theoretically). We need data analysis solutions to overcome our cognitive limitations and transform these huge amounts of data into relevant information that could be used in our decision processes.

Fortunately, there has been great progress in data analysis technologies, by which we understand the whole ecosystem of methods for gathering, storing, processing and presenting data, including algorithms and models from domains like statistics, machine learning, decision support and artificial intelligence. We can use them in practice, since increases in computational power and networking bandwidth made data processing, storing and sharing much more available. As a result, our decision making becomes depended on multiple data processing components and heads towards including AI agents, as frontends to external analysis services, or operating as active participants.

### 2.1.BENEFITS AND RISKS
The benefits of relying our decision processes upon data analysis solutions are undeniable, as the results can be more effective, accurate and rational. Decision-making can use technology in many different ways, from support for data exploration, through acquisition of our preferences, to delegating the whole processes to autonomous systems. Let's take a look at an example of a simple configuration of human decision makers, who are presented with several decision options analyzed by external analysis algorithms and models using data from multiple sources (see Figure 1). This is a common scenario that covers, for example, a company trying to change their operations to be more data driven.

Unfortunately, there are potential problems and risks related to complexity, unverified assumptions and dependencies on external entities in such scenarios. The quality of our decisions depends on multiple key elements. First, we have data, which may contain errors, can be corrupted, altered, outdated or just not available when needed. Secondly, there are analysis methods and services returning results that can be incorrect, of poor quality, or just different from users' needs and expectations. Finally, we have a presentation of results to a decision maker, who needs to correctly interpret them in order to make a decision in a local context. Each of these elements is susceptible to failures, which can result from accidental errors or intentional actions of a 3rd party. In that second case, we can talk about threats against decision-making and data analysis processes.

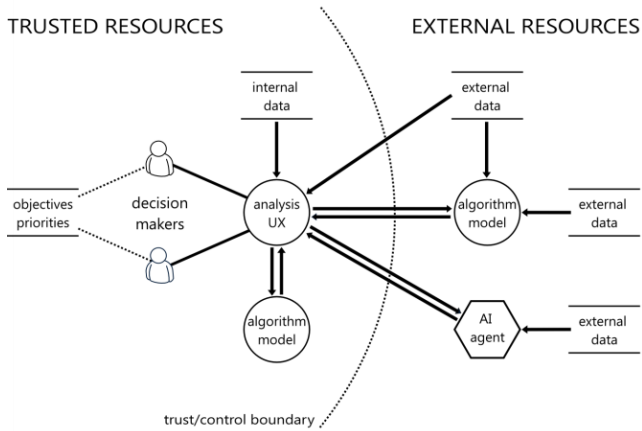

**Figure 1** A generic decision-making process based on data analysis with external data sources and algorithms/models

## 2.2. GOALS AND PRIORITIES

Decision processes based on data analysis solutions should not be considered only on the technical level, as they are usually associated with some real problems and operate in broader contexts, with specific goals and priorities. Let's take a look at an example of a simple scenario with a user interested in purchasing a product. The user searches for the item in an online store and receives a ranking of options to choose from. The analysis process is executed by the store and is external to a user, who has no control or even insight into its details.

The user's goal can be simple - to find a product that is the best balance of price and quality (measured with ratings), but the actual analysis may be conducted with different objectives and include a criterion like state of an inventory. The goals of a user are therefore not fully compatible, or may be even in conflict, with the objectives of the analysis process, upon which he depends. Scenarios with impact on an individual user obviously cannot be ignored, but this becomes a critical problem for decision-making taking place in social contexts.

## 2.3. SOCIAL CONTEXTS

Decision process can involve an individual, a social group, a company, a community or a society. For the purpose of this discussion, let's assume that the social context can mean two things: involvement of multiple participants in a decision-making process, or social impact of multiple persons affected by a decision. In both of these cases, the goals and priorities of a decision process usually need to be clarified, agreed to between multiple parties (often through negotiations) and are critical for success. They cannot be easily updated and any inconsistencies between goals of a decision process and objectives of underlying data analysis must be avoided.

Since decision processes depend on results from data processing components, entities controlling these solutions can indirectly participate in decision-making. If we want our processes to be reliable and trustworthy, we need to define goals and specify functional requirements for data sources and analysis solutions that are about to be used. But for that we need solid understanding of all dependencies, relationships between different components and risks related to specific base technologies. This is usually not easy, even without social elements, due to the complexity and dynamic nature of modern technical systems. Fortunately, this class of problems is not necessarily new or unique.

## 3. THREAT MODELING

Understanding complex systems and their relationships with an environment is a common challenge in information and software security. Even though most security issues are still related to low-level vulnerabilities in code or configuration, we need a big picture to understand existing threats and their consequences. In order to do that, we use threat modeling methodologies to evaluate the design of information processing systems in a security context [1]. We look at a system, its dependencies, assumptions, and external interactions, from the attacker's point of view and try to identify actions that could compromise the system's security properties, like confidentiality, integrity or availability [2].

Such analysis is strongly based on concepts of trust and control - for example, it covers identification of trust boundaries separating our system from external entities, as inbound data flows crossing them require input validation (in our context, outbound data might also require inspection). A resulting threat model can be a base for implementing mitigations against enumerated threats and coordinating overall security efforts, including very specific investments like penetration testing. But most of all, it can help with making informed decisions about design and application of specific technical solutions.

## 3.1. SOCIO-TECHNICAL SYSTEMS

We can apply experience from software threat modeling to decision processes supported by data analysis, but we will need some changes and extensions. The system in our scope reaches beyond technical components and includes additional social and business dimensions, with special emphasis on human elements in some scenarios. We need to extend the scope of analysis to socio-technical systems [3], which are more dynamic, less transparent and harder to model. We also need to revisit some of the base terms to update their definitions. For example, the set of assets that we are trying to protect should include consistency of goals and priorities between a decision process and different underlying data processing components.

Successful attacks against complex systems can be sophisticated, subtle and often remain undetected (what is usually one of the metrics of their success). In socio-technical systems, an attack may have social, economic or political impact, and it may not be obvious who actually benefits from it (phenomenon of fake news is a great example, [4]). The practical requirements for threat modeling socio-technical systems can be based on classical security and privacy methods and practices, but need to go beyond and cover all non-technical elements and relationships between them, which can be very specific to a situation or application.

## 3.2. THREATS

The general threat against decision processes supported by data analysis can be defined as an intentional activity aimed at disrupting these processes or changing their outcome. In most cases, such threats are possible due to dependencies on data sources and analysis components that are outside our control.

Discussion of specific attacks is beyond the scope of this paper, but let's take a look at sample questions indicating possible threats:

a. Data analysis critically depends on the input data, which become more important than algorithms. Do we know if specific data sources are trusted? Are origins of all data flows known and verified? Are the data complete and up-to-date? Are they securely stored and transferred all the time? What transformations have been applied? Do we know the details of data cleaning, normalization, outliers' removal?

b. The second group are algorithms and models, which can be external to our process (often due to their proprietary status). Do we know if results from external algorithms/models are correct? Do we know how models were trained and evaluated? Were data used in training representative for our application? Do they still meet our current requirements and expectations? Do we know what are the objectives and priorities of an AI agent we integrate with?

c. Last but not least, we need to look at threats aimed directly at decision makers, their contexts and cognitive abilities. Are all the pieces of information necessary to make a decision available? Are the results presented in a readable and unequivocal way? With annotations and explanations where needed? Do we always know where specific results are coming from? Can they be easily traced back to source data? Do we understand the impact of specific results on our decision?

### 3.3. MITIGATIONS

Understanding the threats is the first step; in the second one we need to implement proper mitigations addressing these threats. We have established technical mitigations for most of classical threats, related to underlying information processing infrastructure – for example, we know how to store and transfer data securely. Mitigations for new threats however, related to the nature of socio-technical systems, or the context of specific data analysis applications, need to be researched.

Effective mitigations can be technical, but also organizational or legal. The list below contains some examples related to the questions from the previous section:

a. In the scope of data, we have technical capabilities to verify their integrity or origin, but we need clear rules for data ownership, quality metrics, tracking data flows and processing history. Some cases may be difficult, for example when data are contributed by many anonymous users. External data should always be validated, but also outbound data may need checks aimed at verification if no confidential or PII data are shared with external entities. All interactions with data might be also recorded to support nonrepudiation.

b. When it comes to analysis algorithms and models we may need information like up-to-date metrics, configurations or training summaries. The more complex model, the more difficult it is to validate its results in a functional context of an application and we may need independent evaluations, explanations [5] or certifications. Sometimes we should analyze the same data several times using different models and compare results. In some cases, like advanced proprietary solutions or autonomous AI agents, a specific contractual service agreement, covering also analysis objectives and priorities, may be required.

c. Finally, threats against human elements can be addressed by properly designed analysis user experiences, with adjustments to context of specific application and different roles or tasks. Individual needs, requirements and preferences can be managed with advanced personalization and visual decision support focused on identification of possible gaps, ambiguities or inconsistencies. There should always be a path to get from visualization products, through all applied transformations and analysis methods to the original source data.

### 4. CONCLUSION

Decision processes supported by data analysis technologies have great potential, as they can make our decisions more effective, accurate and rational. But for that we need processes and technical solutions that are reliable, trustworthy and cannot be easily compromised. This requires good understanding of our goals and requirements, the assets we want to protect, the related threats and limits of available mitigations.

Threat modeling methodologies adapted to socio-technical systems and the nature of data analysis applications can become a solid foundation for gaining such insight. They can be useful not only for designing and implementing technical mitigations, but also for establishing new rules and best practices for emerging types of data analysis businesses, or shared research efforts, with clearly defined requirements towards creating fair relationships and healthy data ecosystems.

### 4.1. FUTURE

Using threat modeling methodologies is a great starting point, but it is only a first step. Security efforts are continuous in their nature; as technologies and their applications are changing, so are the related threats, what leads to the need for improved mitigations. This seems especially important for decision-making in social contexts, due to complexity, dynamics, and potentially broad impact of these scenarios.

And complexity will increase along with the development of technologies, not only in the scope of data analysis. New types of user interfaces, including augmented or virtual reality, will enable new data analysis experiences, new cooperation scenarios and much closer integration with AI agents. This will also inevitably result in new types of threats against human elements, or decision makers in our specific context.

We need to continuously investigate threats in new data analysis scenarios and applications, and we should use lessons from information and software security whenever possible. We cannot focus only on the benefits and opportunities of new technologies. If we do, we may soon find our decision processes to be very effective and accurate, but not necessarily aligned with our goals and priorities.